\documentstyle[12pt,twoside]{article}
\input{psfig}
\begin{document}
\vspace*{1.1cm}
\begin{center}
{\large
GALAXY  CLASSIFICATION
BY HUMAN EYES AND BY ARTIFICIAL NEURAL NETWORKS
}
\vspace{1.1cm}\\
{
Ofer Lahav  \\
Institute of Astronomy, Madingley Road\\
Cambridge CB3 OHA, UK\\
e-mail: lahav@mail.ast.cam.ac.uk
\vspace{0.57cm}\\
%
%
}
\end{center}

{\small {\bf Abstract --}
The rapid increase in data on galaxy images
at low and high redshift
calls for re-examination of the classification schemes
and for new automatic objective methods.
Here we present a classification method by Artificial Neural Networks.
We also  show results from a comparative study we carried out
using a new  sample of 830 APM digitised galaxy images.
These galaxy images were classified by 6 experts independently.
It is shown that the ANNs can replicate
the classification by a human expert
almost to the same degree of agreement
as that between two human experts,
to within 2 $T$-type units.
Similar methods can be applied to
automatic classification of galaxy spectra.
We illustrate it by Principal Component Analysis of galaxy spectra,
and discuss future large surveys.

\vspace{0.57cm}
}
\begin{center}
{\small {\bf key-words --}
methods: data analysis - galaxies
}
\end{center}
\vspace{1cm}
\noindent 1. {\large INTRODUCTION}
\vspace{0.5cm}

\noindent

The morphological classification of bright galaxies is still
mainly done visually by  dedicated individuals,
in the spirit of Hubble's (1936) original scheme
and its  modifications
(e.g. Morgan 1958, de Vaucouleurs 1959, Sandage 1961,
van den Bergh 1976).
It is remarkable that  these somewhat subjective
classification labels  for galaxies
correlate well with physical properties
such as colour and  dynamical properties.
However, one would like eventually to devise
schemes of classification, which can be related to the physical
processes of galaxy formation.
While  there have  been in recent years
significant advances in
observational techniques (e.g. telescopes, detectors and
reduction algorithms) as well as in  theoretical modelling
(e.g. N-body and hydrodynamics simulations),
galaxy classification remains a subjective area.

Quantifying  galaxy morphology is important for various reasons.
First, it provides important clues to the origin of galaxies and their
formation processes.
For example,
understanding the origin of the type frequency and the
density-morphology relation
is of fundamental importance.
Second, galaxies can also be used as standard candles.
As such they can be used to measure
redshift-independent distances by methods  such as the luminosity-rotation
velocity (Tully-Fisher) relation for spirals and the diameter-velocity
dispersion for ellipticals.
Clearly any observational programme requires an {\it a priori} target list
of objects for photometric or spectrographic measurements.

Therefore galaxy classification is important for both practical
reasons of producing large catalogues for statistical and observational
programs, as well as for establishing
some underlying physics (in analogy with the H-R diagram for stars).
Moreover, understanding the morphology of galaxies at low redshift
is crucial for any meaningful comparison with
galaxy images obtained with the Hubble Space Telescope
at higher redshift ($ z \sim 0.4$).

Most of our current knowledge of  galaxy morphology is based on
the pioneering work of several dedicated observers who
classified thousands of galaxies  and catalogued them.
However, projects  such as the APM and the Sloan digital
sky surveys  yield millions of galaxies.
Classifying very large data sets is obviously  beyond the
capability  of a single person.
Clearly, classification problems in Astronomy  call for new approaches
(e.g. Thonnat 1988;  Odewhan et al. 1991;  Francis et al. 1992;
Spiekermann 1992; Storrie-Lombardi et al. 1992;
Doi et al. 1992; Serra-Ricart et al. 1993; Abraham et al. 1994).

Artificial Neural Networks (ANNs)  have recently been utilised in Astronomy
for a wide range of problems, e.g. from adaptive optics to
galaxy classification
(for review see Miller 1993 and Storrie-Lombardi \& Lahav 1994).
The ANNs approach should be viewed as a general
statistical framework, rather than as an esoteric approach.
Some special cases of ANNs are statistics we are all
familiar with.
However, the ANNs can do better, by allowing non-linearity.
Here we  illustrate these points
by examples from the  problem of morphological  classification
of galaxies, using the ESO-LV (Lauberts \& Valentijn 1989) sample
with 13 parameters and $\sim 5200$ galaxies, as analysed by ANNs
in Storrie-Lombardi et al. (1992) and Lahav et al. (1995),
and for a new sample of $\sim 830 $  APM galaxies (Naim et al. 1994, 1995).

The outline of this review is as follows.
In \S 2 we present a comparative study between experts,
in \S 3 we discuss ANNs and their application to the morphological
classification problem, and in \S 4 we consider spectral classification
of galaxies.

\newpage
\vspace{1cm}
\noindent 2. {\large HUMAN CLASSIFICATION OF  APM GALAXIES }
\vspace{0.5cm}

\noindent

 The motivation for performing  a
comparison between different experts is two-fold.
(i) To study systematically the degree of agreement and reproducibility
between observers.
(ii) To use the human classification as `training sets' for
the Artificial Neural Networks and other automatic classifiers.

We have defined a sample
from the APM Equatorial Catalogue of galaxies
(Raychaudhury et al. 1995)
 selected from IIIaJ (broad blue band) plates taken
with the UK Schmidt telescope at Siding Spring, Australia.
 We chose a subsample of 831  galaxies with
major diameter $D\ge 1.2$ arcmin.
The galaxies were scanned in
raster mode at a resolution of 1 arcsec by
the APM facility at Cambridge.

R. Buta, H. Corwin,  G. de Vaucouleurs, A. Dressler,
J. Huchra and  S. van den Bergh,
(hereafter RB, HC, GV, AD, JH and vdB, respectively)
kindly classified the {\it same} images on the $T$ system
(a conversion to this system was done
in the case of vdB).

\begin{figure}
\caption{Four APM galaxy images and their classification by six
experts and
 and RC3.
The $T$-type classification of
NGC2811 by (RC3, RB, HC, GV, AD,  JH, vdB)
is (1, 1, 1, 1, 1, 1, 1),
of NGC3200 (4.5, 5, 5, 4, 5, 4, 3),
of NGC4902 (3, 3, 4, 3, 3, 5, 3)  and of
NGC3962 (-5, -3, 0, -5, -3, -1, -5).
}
\end{figure}

Four  examples of the human classification are given in Figure 1.
Statistically, all  6 experts agreed on the exact $T$-type
for only
8 galaxies out of the 831
(i.e. less than 1 \%).
Agreement between pairs of observers in excess of 80 \%
are obtained only to within 2 types.
GV and vdB, who classified galaxies over many more years
than the others, were rather conservative
and  did not classify  about a third  of the galaxy images  which
are saturated or of low quality.
The other observers were more liberal and classified almost all the
galaxies.

For each pair of observers $a$ and $b$ the variance was calculated
(cf. Buta et al. 1994):
$$
\sigma_{ab}^2 = {1 \over N_{ab} } \sum_i [ T_{a,i} - T_{b,i} ]^2,
\eqno (1)
$$
where the sum is over
the $N_{ab}$  galaxies for which both observers gave a classification.
The rms dispersion between RC3 (de Vaucouleurs et al. 1991)
and any of the observers
(2.2 $T$-units on average) is larger than between 2 observers who looked
at the {\it same} APM images
(between 1.3 to 2.3  $T$-units,  1.8 on  average).
This reflects the fact that any classification depends on the colour, size
and quality of the images used, i.e. there is no
`universal' classification.
Another interesting result  is that observers who belong to the same
`school' agree better with each other than with others.
For example, the dispersion between deV and HC is only 1.5 and between
HC and RB only 1.3 units. This indicates that
systematic `training' can reduce the scatter
between two human experts.
Detailed analysis and interpretation of this comparison will appear elsewhere
(Naim et al. 1994, Lahav et al. 1994).

As we show below,
it is  encouraging that the  dispersion
we found between the ANN and an expert
is similar to the dispersion between two human experts.

\vspace{1cm}
\noindent 3. {\large  AUTOMATIC  CLASSIFICATION  BY
ARTIFICIAL  NEURAL NETWORKS}
\vspace{0.5cm}

The  challenge is to design a computer algorithm which
will reproduce classification
to the same degree a student or a colleague of the human expert can do it.
Such an automated procedure usually involves two steps:
(i) feature extraction from the digitised image, e.g.
the galaxy profile, the extent of spiral arms, the colour of the galaxy,
or an efficient compression of the image pixels
into a smaller number of coefficients (e.g. Fourier
or Principal Component Analysis).
(ii) A classification procedure,  in which a computer `learns' from
a `training set' for which a human expert provided his or her classification.

Artificial Neural Networks (ANNs), originally suggested as simplified models
of the human brain, are  computer algorithms which provide
a convenient general-purpose framework
for classification (Hertz et al. 1991).
ANNs are related
to other statistical methods common in Astronomy and other fields.
In particular  ANNs  generalise
Bayesian methods, multi-parameter fitting,
Principal Component Analysis (PCA), Wiener filtering
and regularisation methods (e.g. Lahav 1994 for a summary).

\vspace{1cm}
\noindent  {\it 3.1 ANNs as non-linear minimization algorithms }
\vspace{0.5cm}

\noindent
It is very common in Astronomy
 to fit a model with several (or many) free parameters to
the observations. This regression
is usually done by means of $\chi^2$ minimization.
A simple example of a `model' is a polynomial with the coefficients as
the free parameters.
Consider now  the  specific problem of morphological classification
of galaxies. If the type is $T$
(e.g. on de Vaucouleurs' numerical system [-6,11])
and we have a set of  parameters ${\bf x}$ (e.g. diameters and colours)
then we would like to find free parameters ${\bf w}$ (`weights')
such that
$$
\sigma^2  =  {1 \over N_{gal}} \; \sum_i [T_i - f({\bf w}, {\bf x_i})]^2,
\eqno(2)
$$
where the sum is over the galaxies, is minimized.
The function $f({\bf w}, {\bf x})$ is the `network'.
Note the similarity between eq. (2) and eq. (1).
Rather than looking at the variance between two experts,
we minimize here the variance between the expert and the network.
Commonly $f$ is written in terms of
$$
z  = \sum_k  w_k x_k,
\eqno (3)
$$
where the sum here is over the input parameters to each node.
A `linear network' has $f(z)=z$, while a non-linear transfer function
could be a sigmoid $f(z)= 1/[1+ \exp(-z)]$ or $f(z) = \tanh (z)$.
Another element of non-linearity is provided by the `hidden-layers'.
The `hidden layers' allow curved boundaries around clouds of data
points in the parameter space.
While in most  computational problems we only have 10-1000 nodes,
in the brain there are  $\sim 10^{10}$ neurons, each with
$\sim 10^{4}$ connections.


For a given Network architecture the first step is
the `training' of the ANN.
In this step
the weights
are determined by  minimizing `least-squares' (e.g. eq. 2).
The Backpropagation algorithm (Rumelhart, Hinton \& Williams 1986)
is one of the most popular ANN minimization algorithms.
However, there are other more efficient methods such
as Quasi-Newton (e.g. Hertz et al. 1991).

The interpretation of the output depends on the network configuration.
For example, a single output node  provides an `analog' output (e.g.
predicting the type or luminosity of a galaxy), while several output nodes
can be used to assign Bayesian probabilities to different classes
(e.g. 5 morphological types of galaxies).


\vspace{1cm}
\noindent  {\it 3.2 The  Bayesian  connection}
\vspace{0.5cm}

\noindent
A classifier can be formulated
from first principles according to Bayes theorem:
$$
P(T_j|{\bf x}) =  { { P({\bf x} | T_j) \; P(T_j) }
\over { \sum_k P({\bf x} | T_k) \; P(T_k) } }
\eqno (4)
$$
i.e. the {\it a posteriori}
probability for a class $T_j$ given the parameters vector
${\bf x}$
is proportional to the  probability for data
given a class (as can be derived
from a training set) times the {\it prior} probability for a class
(as can be evaluated from the frequency of classes in the training set).
However, applying eq. (4)  requires parameterization of the probabilities
involved. It is common, although not always adequate, to use
multivariate Gaussians.

It can be shown that the ANN behaves like
a Bayesian classifier, i.e. the output nodes
produce  Bayesian {\it a posteriori} probabilities
(e.g. Gish 1990), although it does not implement
Bayes theorem directly.
It is reassuring (and should be used as a diagnostic) that
the sum of the probabilities in an `ideal' network
add up approximately to unity.
For more rigorous and general  Bayesian approaches for modelling  ANNs see
MacKay (1992).

\vspace{1cm}
\noindent {\it 3.3 PCA, data compression and unsupervised algorithms}
\vspace{0.5cm}

\noindent

Principal Component Analysis (PCA) allows
reducing the dimensionality
of the input parameter space.
A pattern can be thought of as being characterized by a point in an
$M$-dimensional parameter space.
One may wish a more compact data description,
where each pattern is described by $M'$ quantities,
with $ M'\ll M$. This can be
accomplished by Principal Component Analysis (PCA), a well known statistical
tool commonly used in Astronomy
(e.g. Murtagh \& Heck 1987 and references therein).
The PCA method is also known in the literature as
Karhunen-Lo\'eve or Hotelling  transform,
and is closely related to the technique of Singular Value Decomposition.
By identifying the {\it linear}
combination of input parameters with maximum variance, PCA
finds $M'$ variables (Principal Components)
that can be most effectively used
to characterize the inputs.

PCA is in fact an example of `unsupervised learning', in which an
algorithm or a linear `network' discovers for itself features and
patterns (see e.g. Hertz et al. 1991 for review).  A simple net
configuration $M:M':M$ (known as encoder)
with linear transfer functions allows finding
$M'$ linear combinations of the original $M$ parameters.  The idea is
to force the output layer to reproduce the input layer, by
least-squares minimization.  If the number of `neck units' $M'$ equals
$M$ then the output will exactly reproduce the input. However, if $M'
< M$, the net will find, after minimization, the optimal linear
combination.  By changing the transfer function from linear to
non-linear (e.g. a sigmoid) one can allow `non-linear PCA'.
Serra-Ricart et al. (1993) have compared standard PCA to non-linear
encoder, illustrating how the latter successfully identifies classes
in the data.

\vspace{1cm}
\noindent  {\it 3.4 Recent results for
galaxy morphological classification by ANNs  }
\vspace{0.5cm}

Storrie-Lombardi et al. (1992)  and  Lahav et al. (1995)
have analysed with ANNs
the ESO-LV (Lauberts \& Valentijn 1989) sample
of about 5200 galaxies, using 13 machine parameters.
Using a network configuration 13:3:1  (with 46 weights, including `bias')
for the ESO-LV galaxy data,
with both the input data and the output $T$-type scaled
to the range [0, 1] and with sigmoid transfer functions,
we get dispersion $\Delta T_{\rm rms}  \sim 2.1$
between the ANN and the experts (LV)
over the $T$-scale [-5, 11].

For a net configuration 13:13:5, where the output layer corresponds
to probabilities for
5 broad classes (E, S0, Sa+Sb, Sc+Sd, Irr),
we found a success rate for perfect match of 64 \%.
Our experiments  indicate that non-linear ANNs
can achieve better classification than the naive Bayesian classifier
with Gaussian probability functions, for which the success rate is only
56 \%.

More recently, we have applied (Naim et al. 1995) the same techniques
to the APM sample of 830 galaxies described above, by extracting
features directly from the images, and training the net on the human
classification from the 6 experts.  When the network is trained and
tested on individual expert, the rms dispersion varies between 1.9 to
2.3 $T$-units over the 6 experts.  A better agreement, 1.8 $T$-units,
is achieved when the ANN is trained and tested on the mean type as
deduced from all available expert classifications. For more details
and the scatter diagram $T(experts)$ vs. $T(ANN)$ see A. Naim in this
volume.  There is a remarkable similarity in the dispersion between
two human experts and that between ANN and experts !  In other words,
our results indicate that the ANNs can replicate the expert's
classification of the APM sample as well as other colleagues or
students of the expert can do.

Similar successful results are reported by S. Odewhan using the
Minnesota galaxy sample in this volume,  and other interesting
computational approaches to classification
are presented by M. Thonnat.

\vspace{1cm}
\noindent 4. {\large SPECTRAL CLASSIFICATION OF GALAXIES}
\vspace{0.5cm}

Galaxy spectra provide another probe of the intrinsic galaxy
properties.  The integrated spectrum of a faint galaxy is an important
measure of its stellar composition as well as its dynamical
properties. Moreover, spectral properties correlate fairly closely
with morphology. Indeed, as the spectra are more directly related to
the underlying astrophysics, they are a more robust classifier for
evolutionary and environmental probes.  Spectra can be obtained to
larger redshifts than ground-based morphologies and, as 1-D datasets,
are easier to analyse. Although the concept of spectral classification
of galaxies dates from Humason (1936) and Morgan \& Mayall (1957), few
uniform data sets are available and most contain only a small number
of galaxies (e.g. Kennicutt 1992).
  Recent spectral analyses for classification were out carried by
Francis et al. (1992) for QSO spectra,
von-Hippel et al. (1994) and Storrie-Lombardi et al. (1994)
for stellar spectra, and in
particular for galaxy spectra  by Sodr\'e \& Cuevas (1994),
Heyl (1994), and Connolly et al.(1994),

Spectral classification is important for several practical and
fundamental reasons.  In order to derive luminosities corrected for
the effects of redshift the $k$-correction (Pence 1976) must be
estimated for each galaxy. The rest-frame spectral
energy distribution is needed, which can  be obtained by matching
the observed spectrum against templates of local galaxies.

The proportion of sources in each class as a function of luminosity
and redshift is of major interest. Apart from its relevance for
environmental and evolutionary studies, new classes of objects may be
discovered as outliers in spectral parameter space.  Furthermore, by
incorporating spectral features with other parameters
(e.g. colour and velocity dispersion) an `H-R diagram for
galaxies' can be examined with possible important implications for
theories of galaxy formation.

To illustrate some of these ideas, we have performed a PCA analysis of
Kennicutt's sample of 55 galaxy spectra over the rest-frame interval
3712-4110 \AA\ including important features such as [OII] 3727 and the
4000 \AA~ break. While the input contains 200 channels, the First
Principal Component accounts for 75\% of the total variance.  Figure 2 shows
the mean spectra, and the first 3 eigen-vectors as a function of
wavelength. We see that the First PC identifies objectively e.g. the
[OII]3727 line.  For each of the galaxies, the projection of the 200
channels on the First PC axis versus the morphological $T$-type is
shown in Figure 3. The correlation confirms that spectral features,
when efficiently extracted, can be used to classify galaxies.
For similar analyses see Sodr\'e \& Cuevas (1994) and
Connolly et al. (1994).
It is
also possible to use a sample for which both the $T$-type and the
spectra are available and to train an ANN to predict $T$-type (or
$k$-correction) from the spectra, similar to the ANN classification of
galaxy images.

These approaches will no doubt be applied to new large surveys such as
the Sloan Digital Sky survey and the 2-degree-field (2dF) 400-fibre
facility at the Anglo-Australian Telescope. In particular, we intend
to carry out automatic classification for 250,000 galaxy spectra,
proposed to be measured with the 2dF by a UK-Australian collaboration.

\begin{figure}
\caption {
Principal Component Analysis applied to 55 galaxy spectra galaxy
spectra (of Kennicutt 1992), evaluated over the range 3712-4110 \AA.
The mean spectra and the first 3 eigen-vectors (together account for
94 \% of the variance) are shown vs.  the wavelength. The PCA was
computed after removing the mean spectra.  The 1st and 3rd components
are shifted by (+0.3) and (-0.4) to clarify the presentation.  The
plot indicated that the Principal Components identify objectively well
known lines, e.g. the 1st PC finds the [OII] 3727 line.  }
\end{figure}

\begin{figure}
\caption {
Projection of the galaxy spectra  on
the First Principal Component axis
(as derived in Figure 2) vs. the morphological $T$-type.
This analysis illustrates that spectral parameters, objectively  extracted,
can be used to classify galaxies.
}
\end{figure}

\newpage

\vspace{1cm}
\noindent 5. {\large Discussion}
\vspace{0.5cm}

It is encouraging that in the problem of morphological classification
of galaxies,  one of the last remaining subjective areas
in Astronomy, ANNs can  replicate
the classification by a human expert
almost to the same degree of agreement
as that between two human experts, to within 2 $T$-units.

The challenge for the future is  to develop efficient methods
for feature extraction and a `unsupervised' algorithms,
combining multi-wavelength information
to define a `new Hubble sequence' without any prior human
classification.

\vspace{0.5cm}

\noindent


{\bf Acknowledgements} -- I grateful to  A. Naim, L. Sodr\'e and M.
Storrie-Lombardi for their contribution to the work presented here, to
R. Buta, H. Corwin, G. de Vaucouleurs, A. Dressler, J. Huchra and S.
van den Bergh for classifying the galaxy images, to S. Raychaudhury,
M. Irwin and D. Lynden-Bell for helping with the APM sample, and to R.
Ellis, S. Folkes, J. Heyl and S. Maddox for discussions on spectral
classification of galaxies and the 2dF.


\end{document}